\documentclass[sigconf]{acmart}

\usepackage{changes}
\usepackage{multirow}
\usepackage{booktabs}
\usepackage{graphicx}
\usepackage{caption}
\usepackage{subcaption}
\usepackage{enumerate}

\AtBeginDocument{%
  \providecommand\BibTeX{{%
    \normalfont B\kern-0.5em{\scshape i\kern-0.25em b}\kern-0.8em\TeX}}}

\setcopyright{acmcopyright}
\copyrightyear{2020}
\acmYear{2020}

\acmConference[SIGIR '20] {43rd International ACM SIGIR Conference on Research and Development in Information Retrieval}{July 25--30, 2020}{Virtual Event, China}
\acmPrice{15.00}
\acmDOI{10.1145/3397271.3401125}
\acmISBN{978-1-4503-8016-4/20/07}

\begin{document}
\fancyhead{}

\title{
A Generic Network Compression Framework for Sequential Recommender Systems
}

\author{Yang Sun}
\affiliation{%
  \institution{University of Science and Technology of China}
 }
\email{yang.sun@siat.ac.cn}

\author{Fajie Yuan}
\authornote{Joint first author.}
\affiliation{%
  \institution{Tencent}}
\email{fajieyuan@tencent.com}

\author{Min Yang}
\authornote{Min Yang is corresponding author. This work
was conducted when Yang Sun was interning at SIAT, Chinese Academy of Sciences.}
\affiliation{%
  \institution{SIAT, Chinese Academy of Sciences}}
\email{min.yang@siat.ac.cn}


\author{Guoao Wei}
\affiliation{%
  \institution{SIAT, Chinese Academy of Sciences}}
\email{ga.wei@siat.ac.cn}

\author{Zhou Zhao}
\affiliation{%
  \institution{Zhejiang University}}
\email{zhaozhou@zju.edu.cn}

\author{Duo Liu}
\affiliation{%
  \institution{SIAT, Chinese Academy of Sciences}}
\email{duo.liu@siat.ac.cn}


\begin{abstract}
Sequential recommender systems (SRS) have become the key technology in capturing user's dynamic interests and generating high-quality recommendations.
Current state-of-the-art  sequential recommender models are typically based on a sandwich-structured deep neural network, where one or more middle (hidden) layers are placed between the input embedding layer and output softmax layer.  
In general, these models require a large number of parameters 
to obtain 
optimal performance. 
Despite the effectiveness, at some point, further increasing model size may be harder for model deployment in resource-constraint devices.
To resolve the issues, we propose a \underline{c}om\underline{p}ressed sequential recommendation framework, termed as CpRec, where  two generic model  shrinking techniques are employed.
Specifically, we first  propose a block-wise adaptive decomposition  to approximate the input and softmax matrices by exploiting the fact that items in SRS obey a long-tailed distribution. 
To reduce the parameters of the middle layers, we introduce  three layer-wise parameter sharing schemes.
We instantiate  CpRec using deep convolutional neural network with dilated kernels given consideration to both recommendation accuracy and efficiency.
 By the extensive ablation studies, we demonstrate that the proposed CpRec can achieve up to 4$\sim$8 times compression rates in real-world SRS datasets. 
Meanwhile, CpRec is faster during training \& inference, and in most cases outperforms its uncompressed counterpart. Our code is available at \url{https://github.com/siat-nlp/CpRec}.

\end{abstract}



\maketitle

\begin{figure}[htbp]
	\centering
	\begin{subfigure}[t]{0.25\textwidth}
		\centering
		\includegraphics[width=1.75in]{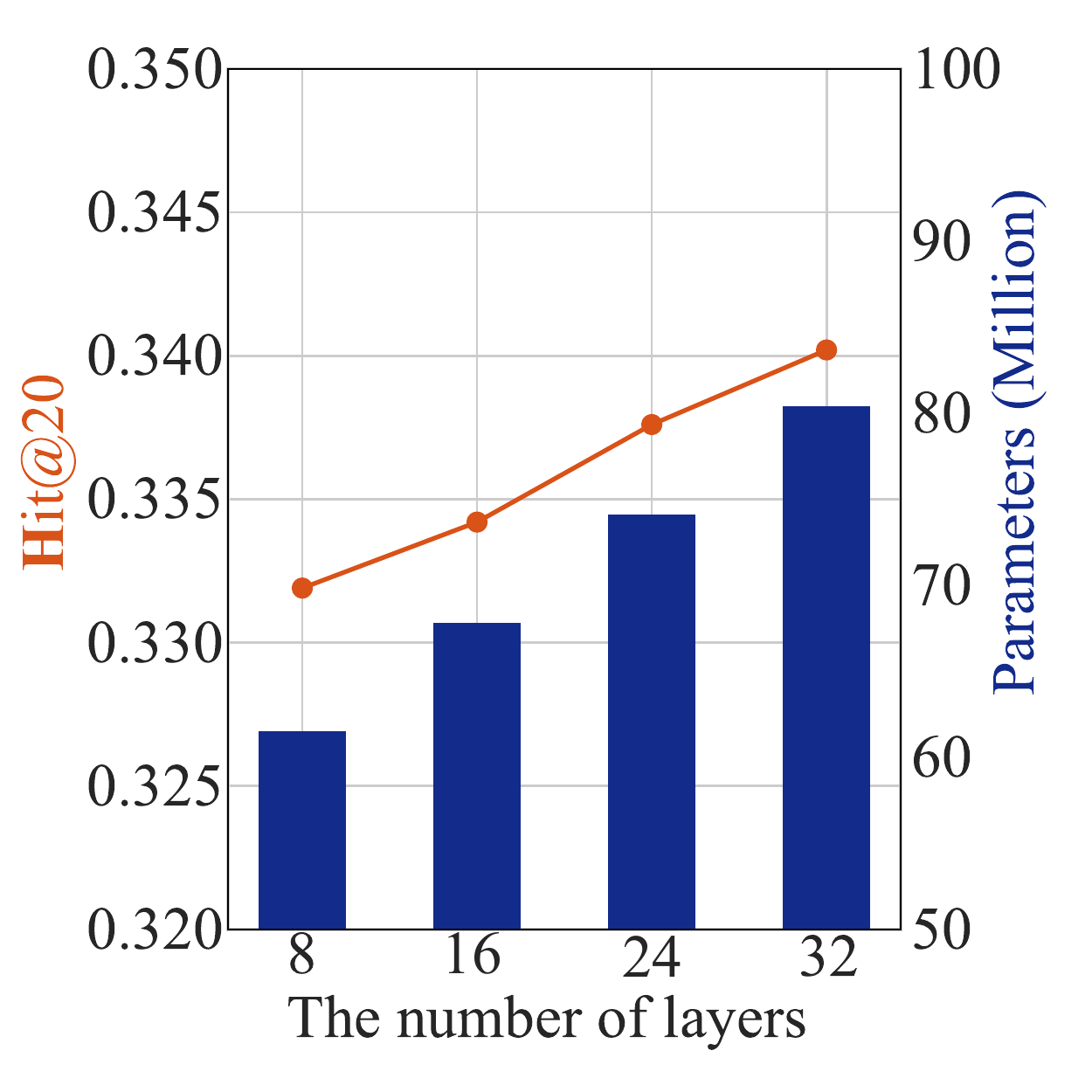}
		\subcaption{Hit@20 vs. the number of layers}
		\label{fig:deep}
	\end{subfigure}%
	\begin{subfigure}[t]{0.25\textwidth}
		\centering
		\includegraphics[width=1.75in]{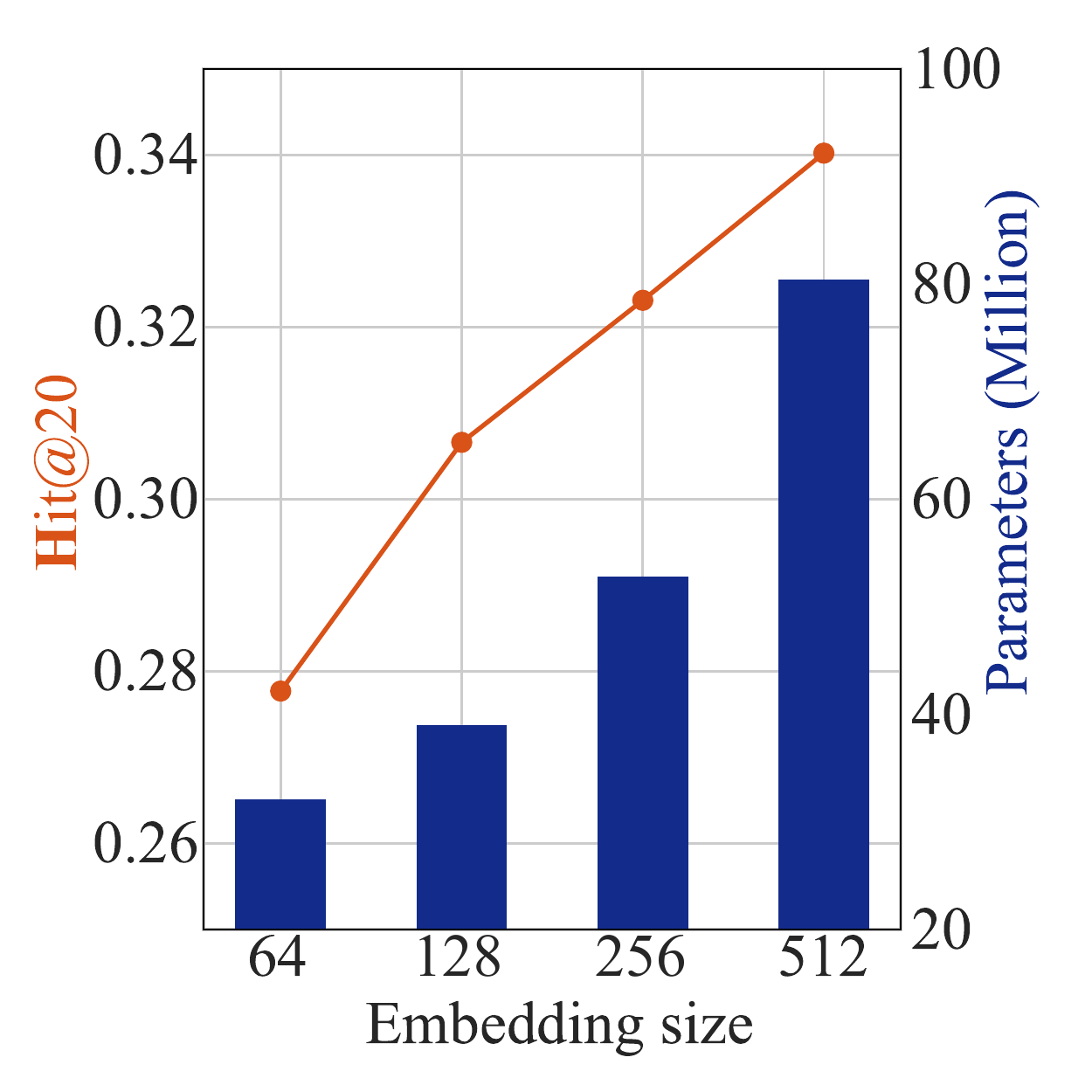}
		\subcaption{Hit@20 vs. embedding size}
		\label{fig:wide}
	\end{subfigure}
	\caption{\small Illustration of prediction accuracy (hit@20) with respect to the model size of NextItNet on ML100 (see Section~\ref{experimentsetup}).  (a)  A deeper model leads to better accuracy where embedding size $d= 512$; (b) A wider embedding leads to highter accuracy where the number of layers  $r$ is set to 32.}
	\label{deepwide}
\end{figure}

\section{Introduction}

Sequential (a.k.a. session-based) recommender systems (SRS) have become a research hotspot in the recommendation field. This is because user interaction behaviors in real-life scenarios often exist in a form of chronological sequences. 
In such scenarios,  traditional RS based on collaborative filtering~\cite{sarwar2001item} or content  features~\cite{lops2011content} fail to model user's dynamic interests and  offer only sub-optimal performance.
By contrast, sequential recommender models based on recurrent neural networks (RNN)~\cite{hidasi2015session,qu2020cmnrec} or  convolutional neural network (CNN) (often with dilated kernels)~\cite{yuan2019simple} have obtained state-of-the-art performance since these models are more powerful in capturing sequential dependencies in the interaction sequence.

In general, modern sequential recommender models based on deep neural networks (DNN) can be split into three major modules: an embedding layer for representing the interaction sequence, a  softmax layer for generating the probability distribution of next item, and one or more hidden (either recurrent or convolutional) layers that are sandwiched between them. To increase the capacity of such models, a larger model size with more parameters is a common method in practice. As shown in Figure~\ref{deepwide},  the prediction accuracy of the sequential recommender model NextItNet~\cite{yuan2019simple} can be largely improved by increasing its model size, i.e.,  using a larger embedding dimension (denoted by $d$) or a deeper network architecture (denoted by $r$). Particularly, NextItNet obtains more than 20\% accuracy gains by simply increasing $d$ from 64 to 512, along with  about 3 times  larger parameters.

Though a large network often brings obvious accuracy improvements, it may also become the major obstacle for model deployment and real-time prediction, especially for memory-limited devices, such as GPU/TPU or end-user devices
Another drawback is that both the training and inference speeds will be impacted by employing larger matrices and deeper networks. 
In addition, many published works also evidence that at some point further increasing model size may  cause the overfitting problem~\cite{yuan2020parameter} or unexpected model degradation~\cite{lan2019albert}. 
Hence, we argue that model compression is essential to achieve compact recommender models that enable real-time responses and better generalization ability.

In fact, the model size problem in the recommender systems domain is much more challenging than other domains, such as  computer vision (CV)~\cite{gong2014compressing} and natural language processing (NLP)~\cite{devlin2018bert,baevski2018adaptive}. For example, in CV, the well-known ResNet-101~\cite{zagoruyko2016wide} trained for ImageNet has only 44.5 million parameters~\cite{boulch2017sharesnet}. One of the largest NLP model BERT-Large (24 layers, 16 attention heads) has around 340 million trainable parameters~\cite{devlin2018bert}. By contrast, in industrial recommender systems, such as  Youtube and Amazon, there exist several hundred million items.
If we simply assume the number of items as 100 million and set the embedding dimension $d$ to 1024, we can achieve over 2$\times$100 billion trainable parameters w.r.t.  embedding \& softmax matrices, which is more than 4000  and 400 times larger than ResNet-101 and BERT-Large, respectively. 
On the other hand,  parameters from the middle layers cannot be ignored in medium-sized or small-scale recommender systems, such as the future in-vehicle recommender systems~\cite{luettin2019future}, where memory consumption may be dominant by both middle layers and  embedding matrices. Taking NextItNet as an example, the number of parameters in middle layers  is $1024 \times1024\times 3 \times 32 \approx 100$ million (still much larger than ResNet-101) where  3 and 32 is the kernel size and layers, respectively. Actually, in practice, more convolutional layers may be required for better accuracy if user behavior sequences  are longer. Therefore, to reduce the size of sequential recommender models, we need to consider parameter-reduction for both embedding \& softmax matrices and middle layers.
To address the aforementioned issues, we present two generic model compression
 methods to lower memory consumption for SRS. First, to reduce parameters in the
 embedding and softmax matrices  mentioned above, we propose  block-wise adaptive  decomposition to approximate the original large emebdding\footnote{\scriptsize Without special mention, both the input embedding matrix and the output softmax matrix are  unifiedly called an embedding matrix in the following description.} matrices. Specifically, we  separate all candidate items into clusters according to their frequencies, and the embedding matrix of each cluster, referred to as block,  is decomposed by two low-rank matrices, where the rank value  is also determined by the item frequencies in the cluster --- a larger rank value is assigned to the blocks with more frequent  items, and vice versa.
 Our idea here is motivated by the well-known finding that the item frequency distribution in recommender systems is generally long-tailed --- i.e., only a few  items may contain rich information due to their high frequency, while others  may only contain limited information.  Given this structure, a fixed large embedding dimension for all items is redundant and may lead to sub-optimal performance.  By the block-wise adaptive decomposition, we are able to assign different dimensions to the block of each cluster. 
  Second, motivated by the cross-layer parameter sharing method of {ALBERT} \cite{lan2019albert}, we introduce cross-block, adjacent-layer and adjacent-block parameter sharing methods to reduce parameters in the middle layers. Since the two parameter-reduction methods are orthogonal, we can naturally combine them together to achieve a higher compression rate. We name the proposed joint compression framework CpRec. 

We summarize our main contributions as follows.
\begin{itemize}
\item We propose a block-wise adaptive decomposition method to approximate the original large input/output embedding matrices in SRS.
Unlike the standard low-rank decomposition~\cite{denil2013predicting}, our method enables a better and fine-grained approximation by exploiting the inherent structure of item distribution.
To the best of our knowledge, CpRec is the first model compression method in SRS, which directly targets reducing parameter size of the embedding matrices.
 
\item Inspired by the class-based softmax \cite{le2011structured} for language model, we design a probability approximation method based on a tree representation in the softmax layer, which replaces the vanilla softmax and notably reduces the training time. Unlike~\cite{le2011structured}, we perform class-based softmax on the decomposed embedding matrix.
\item We propose three different layer-wise parameter sharing methods to reduce redundant parameters in the middle layers, which effectively constrains the parameter size as the model grows deeper. 
\item  We obtain a compression ratio of 4\textasciitilde8x on four real-world SRS datasets. 
Moreover,  we demonstrate that CpRec outperforms the uncompressed counterpart in most cases and is faster for both training and generating.


\end{itemize}

\section{Related Work}
\subsection{DNN-based SRS}

Recently, deep neural networks (DNNs) have brought great improvements for SRS and almost dominate this field. Thus far,  three types of DNN models have been explored for SRS. Among them, Recurrent Neural Networks (RNNs) are often a natural choice for modeling sequence data~\cite{guo2019dynamic}. GRU4Rec\cite{hidasi2015session,tan2016improved} is regarded as the seminal work that firstly applied gated recurrent units (GRU) architecture for sequential recommendation tasks. Inspired by them, 
a variety of RNN variants have been proposed to address the sequential recommendation problems, such as personalized  SRS with hierarchical RNN~\cite{ying2018sequential}, content- \& context-based SRS~\cite{gu2016learning,smirnova2017contextual}, data augmentation-based SRS~\cite{tan2016improved}.
While effective, these  RNN-based  models seriously depend on the hidden state of the entire past, which cannot take full advantage of  modern parallel processing resources~\cite{yuan2019simple}, such as GPU/TPU. By contrast,  convolutional neural networks (CNNs) and pure attention-based models do not have such limitations since the entire sequence is already available during training. 
In addition, CNN and attention-based sequential models can perform better than RNN recommenders since much more hidden layers can be stacked by the residual block architecture~\cite{he2016deep}. To be more specific, \cite{yuan2019simple}  
proposed a CNN-based generative model called NextItNet, which employs a stack of dilated convolutional layers to increase the  receptive field when modeling long-range sequences.   
Likewise, self-attention based models, such as SASRec~\cite{kang2018self} and BERT4Rec~\cite{ sun2019bert4rec} also obtained  competitive results. Compared with NextItNet, the self-attention mechanism is computationally more expensive since calculating self-attention of all timesteps requires quadratic complexity and memory. 

All above mentioned sequential recommender models  consist of three major modules: two embedding layers for input \& output items and several middle layers sandwiched between them. In this paper, we focus on exploring model compression technology for these sandwich-like  recommender models. For the below description,
we specify CpRec by using  the NextItNet architecture, although it  can be directly applied to a broad range of recommendation models,
such as GRU4Rec and SASRec, etc.

\begin{figure*}[htbp]
    \centering
    \begin{subfigure}[t]{0.23\linewidth}
    \centering
            \includegraphics[width=1.42in]{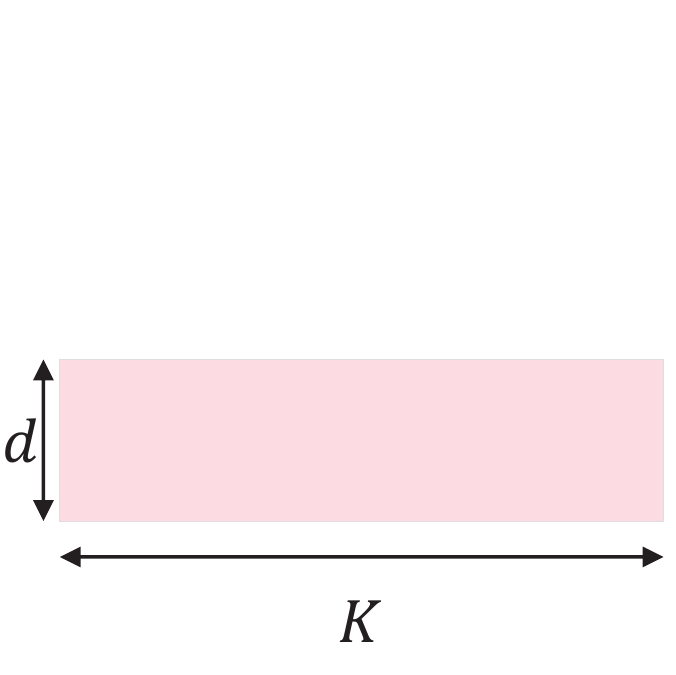}
            \subcaption{Base}
            \label{fig:base}
    \end{subfigure}%
    \begin{subfigure}[t]{0.23\linewidth}
    \centering
            \includegraphics[width=1.55in]{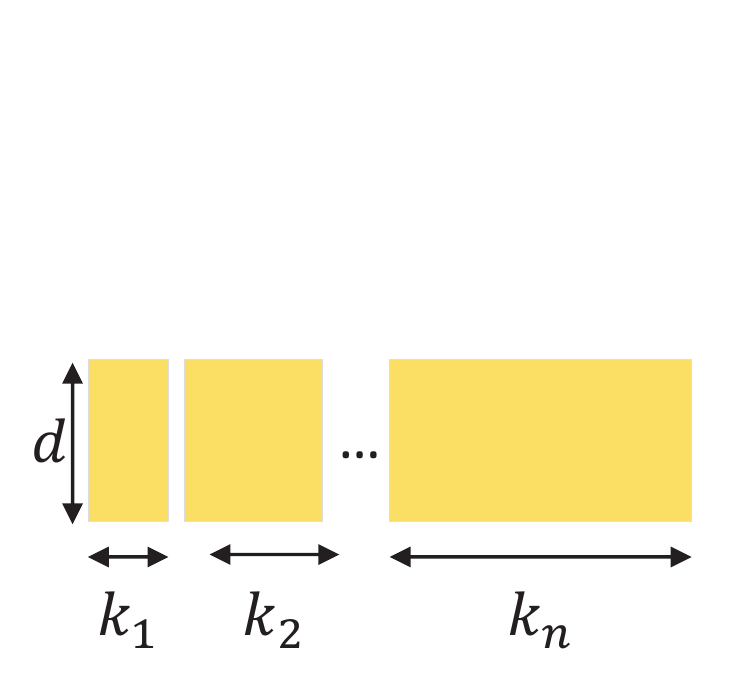}
            \subcaption{Block-partitioning}
            \label{fig:partition}
    \end{subfigure}
    \begin{subfigure}[t]{0.25\linewidth}
    \centering
            \includegraphics[width=1.9in]{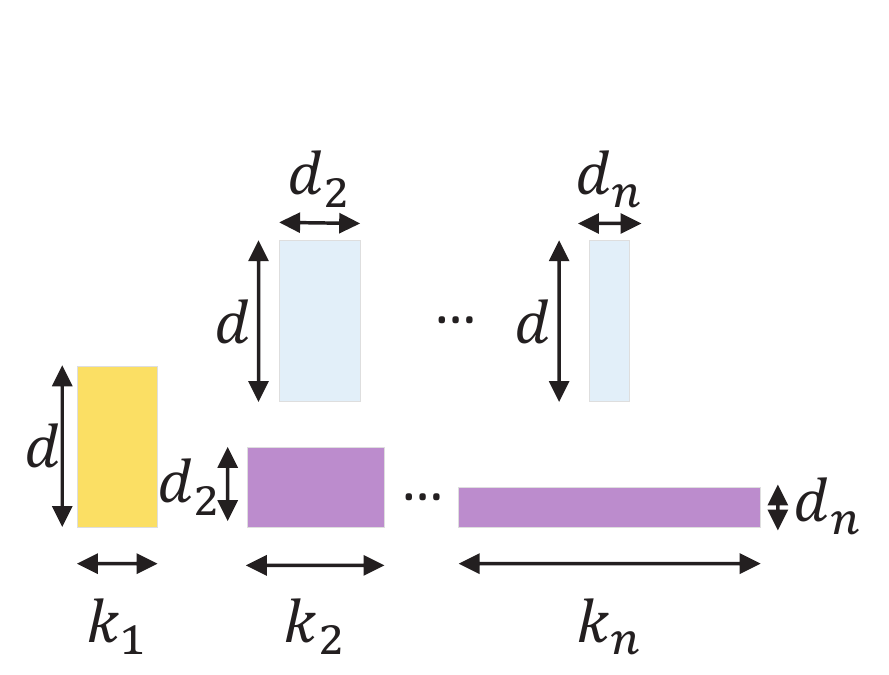}
            \subcaption{\centering Block-wise adaptive decompostion (for input layer)}
            \label{fig:blockinput}
    \end{subfigure}
    \begin{subfigure}[t]{0.25\linewidth}
    \centering
            \includegraphics[width=1.59in]{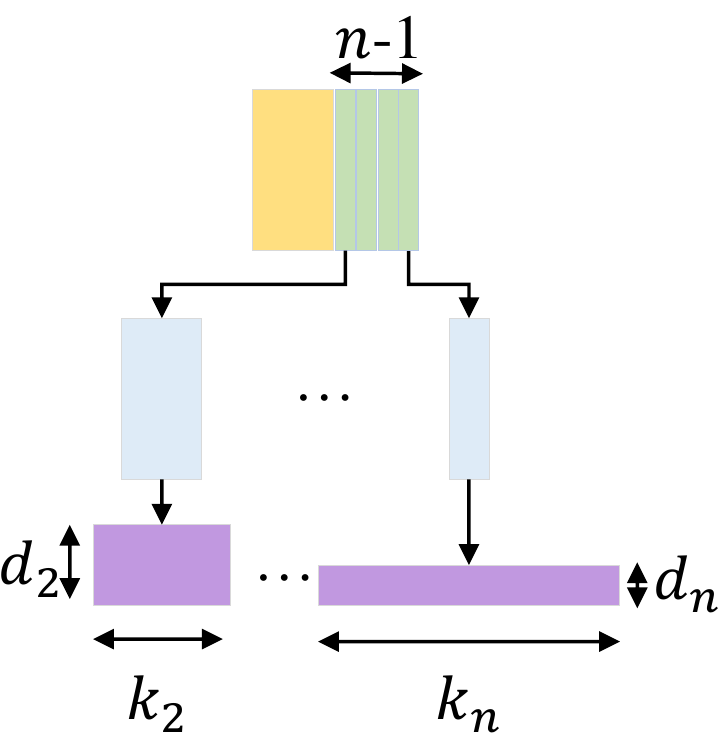}
            \subcaption{\centering  Block-wise adaptive decompostion (for softmax layer)}
            \label{fig:blockoutput}
    \end{subfigure}
    \caption{\small Illustration of the proposed block-wise embedding decomposition methods: (a)  the vanilla input or output embedding  matrix; (b) the embedding matrix of (a) is partitioned into several blocks based on item frequencies; (c) performing low-rank decompostion for blocks in the input layer (d) performing low-rank decompostion for blocks in the output layer based on a two-layer tree structure, where n-1 represents the number of parent classes that the leaf nodes belong to.}
    \label{fig:2}
    \vspace{-0.1in}

\end{figure*}

\subsection{Model Compression}
Training larger and deeper neural networks have become a common practice to obtain state-of-the-art results for a variety of tasks. Meanwhile, the performance gain is usually accompanied by the price of slow processing speed and a huge amount of memory, which makes these models difficult to be deployed on devices with limited resources. 
As such, more and more  attention has been paid to model compression methods. 
In general,  compression methods can be divided into four categories.

 \paragraph{Low-rank factorization} 
 The basic idea of this method is to factorize the original large matrix into a product of two low-rank matrices\cite{yuan2018fbgd}. In terms of model compression,	it is mainly used for compressing fully-connected and convolutional layers. 
	 Specifically, \cite{denil2013predicting, zagoruyko2016wide} argued that there was significant redundancy in the parameterization of deep neural networks. As a result, they 
	used two smaller matrices to represent the weight matrix learned from images to reduce free parameters.
	 \cite{sainath2013low} proposed using low-rank factorization to reduce parameters in the softmax layer.  A more recent work in {ALBERT} \cite{lan2019albert} performed standard low-rank approximation to decompose both the input and softmax layers into two smaller matrices. 
	By contrast, our proposed method is distinct from them because we perform embedding low-rank factorization adaptively for different blocks according to their item frequencies.

\paragraph{Weight quantization} Network quantization uses low-precision float to represent weight values, which is a general compression method for any deep learning model. For instance, \cite{gong2014compressing,wu2016quantized,Han2015arXiv} quantized weight and then applied k-means clustering to identify shared weights.
	  \cite{vanhoucke2011improving} used 8-bit quantization to convert activation and intermediate layer weights, which  
	  easily leads to significant memory compression. Similarly, \cite{xu2018deep} proposed single-level quantization method for high-bit quantization and multi-level method for low-bit quantization. 
	    Due to its generality, network quantization can also be applied to CpRec for further compression.
	  
\paragraph{Parameter pruning and sharing} Parameter pruning improves parameter efficiency by cutting out redundant parameters. For example, \cite{han2015deep, srinivas2015data} pruned unimportant connections iteratively to lower the storage and computation resources required by the model.
	 {ALBERT} \cite{lan2019albert} and \cite{2018arXiv180703819D} adopted the cross-layer parameter sharing method. While model sizes were shrunk notably, the performance of their models also significantly decreased. \cite{baevski2018adaptive} tie the weights of the input embedding layer and output softmax layer to achieve further compression after the low-rank factorization.
	 
\paragraph{Knowledge Distillation (KD)} The basic idea of KD-based compression is to
transfer knowledge from a large, pre-trained teacher model to a student one that is typically smaller.
 Recently, \cite{tang2018ranking} proposed the first KD-based  model for the learning to rank problem in recommendation. They showed that the proposed KD method achieved similar ranking performance but with smaller model size and higher inference efficiency.  However, throughout the paper, they only  investigated the proposed ranking distillation on two very shallow embedding models. The performance of KD on larger \& deeper recommender models keeps completely unknown. In addition, they only obtained  2\textasciitilde2.5x compression rate, which is much smaller than that of CpRec.

To the best of our knowledge, model compression techniques have not been well studied  in recommender systems.
 One reason may be  that authors in existing literature tended to apply a very small embedding dimension (e.g., $d=5$ for IRGAN on Movielens~\cite{wang2017irgan}, $d=10$ for Fossil on Foursquare~\cite{he2016fusing},  $d=20$ \& $30$ for N-MF~\cite{hu2014your} \& LambdaFM~\cite{yuan2016lambdafm} on Yelp, $d=50$ for NARM on YOOCHOOSE1~\cite{li2017neural}
 ) for research purpose. 
The other reason is that thus far there seems no existing literature using deep learning models higher than 20 layers for the recommendation task. However, as clearly evidenced  in Figure~\ref{deepwide}, a large NextItNet with $d=512$ and $r=32$ indeed performs better on the benchmark dataset. 

\section{Methods}

In this section, we present two main model compression techniques to improve the parameter efficiency of SRS. In what follows, we describe the proposed CpRec by using the NextItNet architecture. 

\subsection{Block-wise Adaptive Decomposition}

In recommender systems, a well-known observation is that frequencies of items generally obey a long-tailed distribution \cite{yuan2019simple,yuan2019simple,tu2015activity}, where some ``head'' (or popular) items have a large number of user interactions, yet only a few interactions are available for the ``tail''  items. In view of this, we argue that assigning a fixed embedding dimension to all items is sub-optimal and unnecessary. Intuitively, items with higher frequencies may contain more information than the rare ones, and thus should be assigned with more capacity during training. In other words, the embedding dimensions of more frequent (or popular) items are supposed to be  larger than those of unpopular items. 

An obvious difficulty is that if we set adaptive (i.e.,variable-sized) embeddings to items,  they cannot be directly trained by the typical sequential recommender model due to inconsistent dimensions of middle layers.  To this end, we perform dimension transformation by multiplying a projection matrix. From a reverse perspective, the transformation process is equivalent to a low-rank factorization given that the original large embedding matrix is reconstructed by two smaller matrices. 

More specifically, we  first sort all items based on their frequencies  $S=\{x_1,x_2, \ldots, x_K\}$, where $x_1$ and $x_K$ are the most popular and unpopular items, respectively.  Denote the number of clusters as $n$.  We partition the item set $S$ into $n$ clusters:  $S=S_{1} \cup S_{2}, \ldots, \cup S_{n},$ $S_{1}=\{x_1,x_2, \ldots, x_{k_1}\}$, $S_{2}=\{x_{k_1+1},$ $\ldots, x_{k_1+k_2}\}, $
$ \ldots, S_{n}=$
$\left\{x_{k_{1}+\ldots+k_{n-1}+1}, \ldots, x_K\right\}$, where $S_{\alpha} \cap S_{\beta}=\emptyset$,  $\alpha \neq \beta$, and the number of items of each cluster is $k_{1}, k_{2}, \ldots, k_{n}$, $\sum_{i=1}^{n} k_{i}=K$. 
Accordingly, we can partition the embedding matrix  $E \in R^{K \times d}$ in Figure \ref{fig:base} into blocks $E^{1} \in \mathbb{R}^{k_{1} \times d}, E^{2} \in \mathbb{R}^{k_{2} \times d}, \ldots, E^{n} \in \mathbb{R}^{k_{n} \times d}$ in Figure \ref{fig:partition}, where $d$ is the embedding size. Following similar strategy, the softmax matrix $P \in R^{d \times K}$ of output layer can be  partitioned into blocks $P^{1} \in \mathbb{R}^{d \times k_{1}}, P^{2} \in \mathbb{R}^{d \times k_{2}}, \ldots, P^{n} \in \mathbb{R}^{d \times k_{n}}$. In the following, we will describe block-wise embedding decomposition  for the input embedding matrix and output softmax matrix separately.
\begin{figure*}[htbp]
	\centering
	\begin{subfigure}[t]{0.22\linewidth}
		\centering
		\includegraphics[width=0.815in]{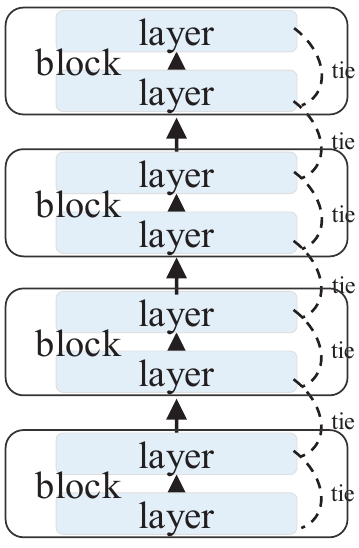}
		\subcaption{Cross-layer}
		\label{fig:crosslayer}
	\end{subfigure}%
\begin{subfigure}[t]{0.22\linewidth}
	\centering
	\includegraphics[width=1in]{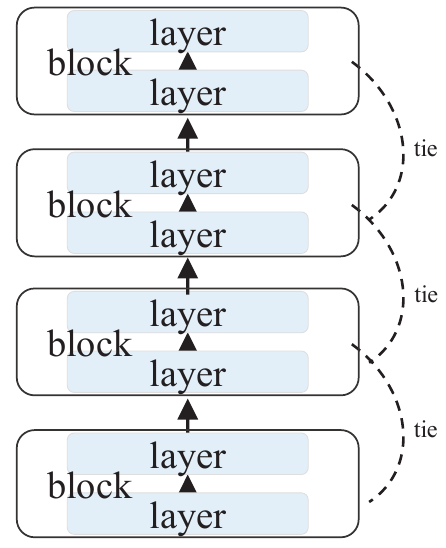}
	\subcaption{Cross-block}
	\label{fig:crossblock}
\end{subfigure}%
	\begin{subfigure}[t]{0.22\linewidth}
		\centering
		\includegraphics[width=0.84in]{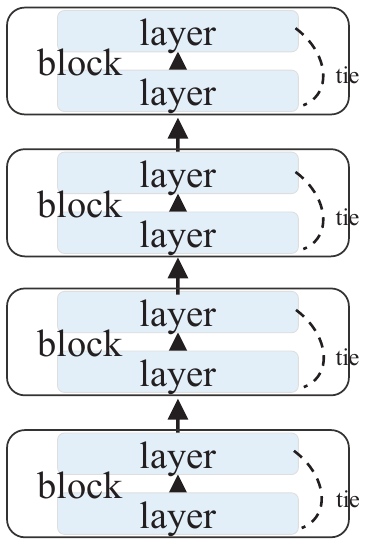}
		\subcaption{Adjacent-layer}
		\label{fig:adjacentlayer}
	\end{subfigure}
	\begin{subfigure}[t]{0.22\linewidth}
		\centering
		\includegraphics[width=0.95in]{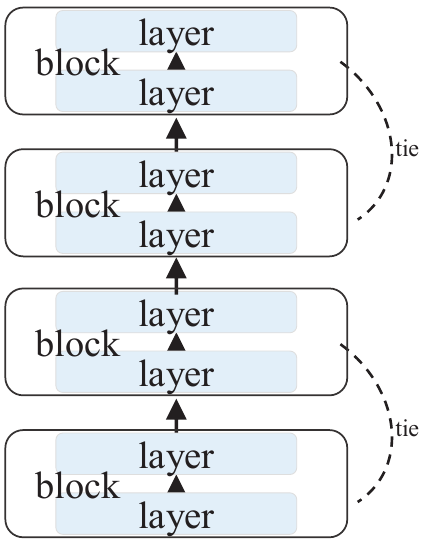}
		\subcaption{Adjacent-block}
		\label{fig:adjacentblock}
	\end{subfigure}
	\caption{\small Illustration of the cross-layer/block parameter sharing and our proposed adjacent-layer/block parameter sharing.}
	\label{fig:layerwise}
	\vspace{-0.1in}
\end{figure*}
\subsubsection{Adaptive decomposition for the input layer}
In the input layer, we factorize the block matrix $E^{j}, j=2,\ldots,n$ with two low-rank matrices $E^{j}=\widehat{E}^{j} W^{j}$, where $\widehat{E}^{j} \in \mathbb{R}^{k_{j} \times d_{j}}, W^{j} \in \mathbb{R}^{d_{j} \times d}$. $d_{j}$ is the factorized  dimension (a.k.a. rank) for the $j$-th  cluster and we decrease it as the cluster index increases since popular items should have higher expressive capacity. Correspondingly, the embedding  representation of each item is different from the one by vanilla look-up operation. Given an item label ID $x$, we use the following equation to denote its embedding vector $v_x \in \mathbb{R}^{d}$:
\begin{equation}
\setlength{\abovedisplayskip}{3pt}
\setlength{\belowdisplayskip}{3pt}
v_x =\left\{\begin{array}{ll}
{E_{x}^{1}} & {\text { if } x \in S_{1}} \\
{\widehat{E}^j_{g} W^{j}} & {\text { if } x \in S_{j} \text { and } j \neq 1}
\end{array}\right.
\end{equation}
where $\widehat{E}^j_g$ represents an embedding vector in the $g$-th row of the $j$-th block, where $g=x-\sum_{i=1}^{j-1}k_i$.  
Through this factorization, we reduce the parameters of input layer from $O(K \times d)$ to $O(k_{1} \times d+\sum_{i=2}^{n}\left(k_{i}+d\right) \times d_{i})$. When $d_{i} \ll d_{1}$, the parameters of input layer would be significantly reduced. Figure \ref{fig:blockinput} illustrates the factorization process.

\subsubsection{Adaptive decomposition for the softmax layer}

It is non-trivial to directly apply the same strategy for the softmax layer since it is unknown which cluster the item belongs to during inference. To address this issue, we  explore a straightforward way by simply softmaxing all prediction scores (i.e., logits) of all clusters and recommending top-N items with N highest probabilities. However, we find that this trivial implementation brings a significant decrease on model performance (around 5\% to 30\% performance drop). We suspect the reason is that  it is not accurate  to project logits calculated from different clusters to the same softmax space.


Inspired by class-based softmax \cite{le2011structured},  we structure these blocks by a two-layer tree, where each tree node represents a single cluster. The paradigm of block-wise embedding for the softmax layer is shown in Figure \ref{fig:blockoutput}. The embedding matrix of the first cluster (i.e., the first node with dimension $d$) is saved to the root node, and other blocks are saved to the leaf nodes in the second layer. For the first cluster, each item is represented by one distinct class; whereas for other clusters, we assign two nodes to each item: a root node using its cluster position as its parent class and a leaf node representing the specific position in this cluster. By doing so, the items in the same cluster share the same parent class.
To be more specific, we use the similar clustering configuration of block-wise embedding in the input layer. A major difference here is that the first block matrix  is extended to $\widehat{P}^{1} \in R^{d \times\left(k_{1}+n-1\right)}$ in the output layer, where $n-1$ represents the number of parent classes that the leaf nodes belong to. The label set of the first cluster is extended to $S_{1}^{\prime}=\left\{1,2, \ldots, k_{1}+n-1\right\}$ , where $k_1+1$ to $k_{1}+n-1$ correspond to the parent class labels of the clusters 2 to n. The other block matrices in the output layer is ${P^{j}}= \widehat{W}^{j} \widehat{P}^{j}$, where $j \in \{2, \ldots, n\}$,  $\widehat{P}^{j} \in \mathbb{R}^{d_{j} \times k_{j}}, \widehat{W}^{j} \in \mathbb{R}^{d \times d_{j}}$. Compared with the vanilla softmax layer, we reduce the number of parameters of the output layer from $O(K \times d)$ to $O((k_{1}+n-1) \times d+\sum_{i=2}^{n}(k_{i}+d) \times d_{i})$. In the rest of this section, we describe how to formulate the objective function during training and how to perform generating during inference in detail. 

During training, to predict the next item given a context vector $h \in \mathbb{R}^d$ (i.e., the final hidden vector of a sequential recommender model), we need to first determine the search space based on the  label of next item, e.g., $x$. If $x$ belongs to the first cluster, we only  compute its logits in this cluster. If $x$ belongs to the other clusters, then we compute the logit in both its parent class and the current cluster. The logit $\hat{y}$ is given as

\begin{equation}
\setlength{\abovedisplayskip}{3pt}
\setlength{\belowdisplayskip}{3pt}
\label{logit}
{\hat{y}=\left\{\begin{array}{ll}
	{h \widehat{P}^{1}} & {\text { if } x \in S_{1} \text { or } \exists  c(x)} \\
	{h \widehat{W}^{j} \widehat{P}^{j}} & {\text { if } x \in S_{j} \text { and } j \neq 1}
	\end{array}\right.} \\
\end{equation}
where we stipulate that each item $x$ belonging to the leaf node have a parent class label c(x) in the first cluster.  Correspondingly, the training process includes two steps. In the first step, the logits of the first cluster are computed, which takes $O(k_1+n-1)$ time. In the second stage, if the item label $x$ belongs to one of the leaf class, we compute the logits of that leaf cluster, which takes $O(k_j)$ time. By doing so,  we reduce the training time from $O(K)$ using vanilla softmax to between $O(k_1+n-1)$ and $O(k_1+k_j+n-1)$ using the block-wise embedding.

Let $\hat{p}$ be the normalized value of $\hat{y}$ by the softmax function. The  loss function $f$ with respect to $\hat{p}$ and ground-truth label vector $y$ is given
\begin{equation}
\setlength{\abovedisplayskip}{3pt}
\setlength{\belowdisplayskip}{3pt}
f(y, \hat{p})=\left\{\begin{array}{ll}
{-\sum_{i=1}^{k_{1}+n-1} y_{i} \log \hat{p}_{i}} & {\text { if } x \in S_{1}} \\
{-\sum_{i=1}^{k_{1}+n-1} y_{i} \log \hat{p}_{i}} {-\sum_{i=1}^{k_{j}} y_{i} \log \hat{p}_{i}} & {\text { if } x \in S_{j} \text { and } j \neq 1}
\end{array}\right.
\end{equation}
Different from the training phase, it is unknown which cluster the item belongs to during inference. Yet, we are able to calculate the probability distributions of items in all clusters according to the condition distribution, given as follows:
\begin{equation}
\setlength{\abovedisplayskip}{3pt}
\setlength{\belowdisplayskip}{3pt}
\label{equ:probability}
p\left(x\right)=\left\{\begin{array}{ll}
	{p\left(x | h, S_{1}\right)} & {\text { if } x \in S_{1}} \\
	{p\left(x | c(x), h\right) p\left(c(x) | h, S_{1}\right)} & {\text { if } x \in S_{j} \text { and } j \neq 1}
\end{array}\right.
\end{equation}
where $p\left(x | h, S_{1}\right) $, $p\left(c(x) | h, S_{1}\right)$ and $p\left(x | c(x), h\right)$  can  all be calculated by Eq.(\ref{logit}). Finally, we are able to recommend the top-N items based on $p\left(x\right)$. In practice, it is usually not necessary to compute the softmax probabilities for all items during the inference phase. We can perform an early-stop search to speed up the generating process. Specifically, if the top-N probability scores are in the first cluster, we do not need to compute scores in the other clusters (i.e., $p\left(x | c(x), h\right)$) since  $p\left(x | c(x), h\right) p\left(c(x) | h, S_{1}\right)$ (where $p\left(c(x) | h, S_{1}\right)<1$) is always smaller than 
the top-N scores of the first cluster.


\begin{figure*}[htbp]
	\centering
	\begin{subfigure}[t]{1\linewidth}
		\centering
		\includegraphics[width=7in]{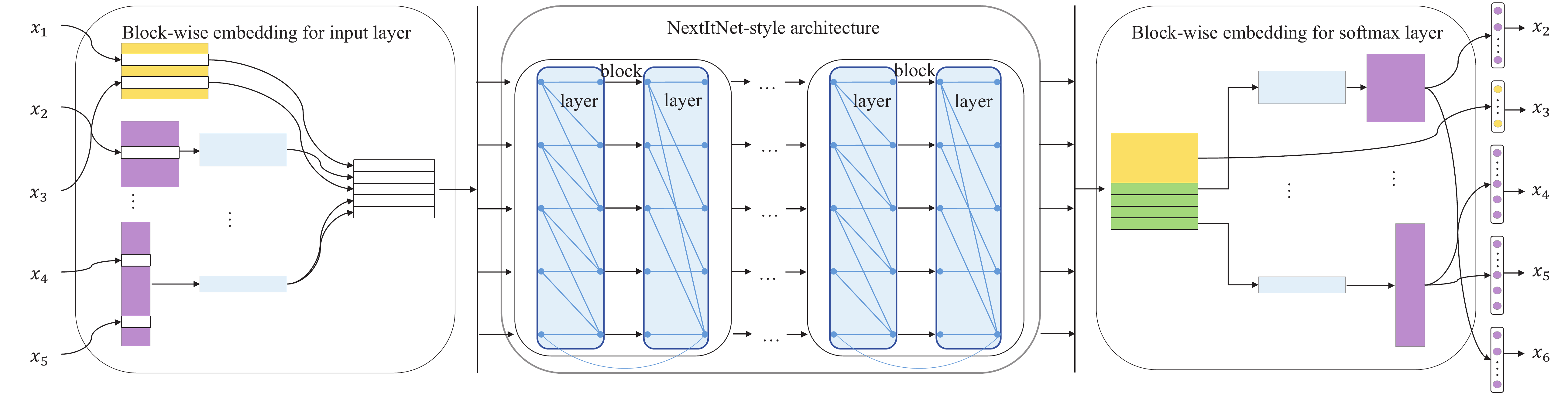}
	\end{subfigure}%
	\caption{\small The neural architecture of CpRec. }
	\label{fig:CpRec}
	\vspace{-0.1in}
\end{figure*}

\subsection{Layer-wise parameter sharing}
User behavior sequence can be very long in many real-world recommender systems, such as short video and news recommendation. To model long-range interaction sequence, a common approach is to build a deeper network architecture. As explained in the introduction section, parameter size in  middle layers may  dominate the overall memory consumption, especially for small-scale applications in mobile or end-user devices. Hence, the proposed compression method in this section is mainly concentrated on sequential recommender models which have a number of middle layers and every two of them are connected by a residual  block (one the most popular ResNet~\cite{he2016deep} structure),
such as NextItNet.

To lower parameter consumption in middle layers, {ALBERT}~\cite{lan2019albert} proposed the cross-layer parameter sharing  approach.
While a large number of redundant parameters are reduced, the expressivity of neural network models is also restricted to some extent. In fact, we notice that the performance significantly decreased in both the original paper and our recommendation task by using the cross-layer sharing scheme. To further evidence this observation, we
propose an advanced version, namely the cross-block parameter sharing as shown in Figure~\ref{fig:crossblock}, where all higher layers reuse  parameters of the first residual block (i.e., the  two bottom layers).


In order to fully utilize the advantage of stacking layers and meanwhile improve parameter efficiency, we propose another two  layer-wise parameter sharing methods: adjacent-layer and adjacent-block parameter sharing. 
 Specifically, the adjacent-layer parameter sharing denotes that the two individual layers in each residual block share the same set of parameters, as shown in Figure~\ref{fig:adjacentlayer}.
 The adjacent-block parameter sharing denotes that parameters are shared between each adjacent residual block, as shown in Figure~\ref{fig:adjacentblock}. Our parameter sharing strategies are considered to bring two major benefits: (1) as a way of regularization, they stabilize the training process and improves model's generalization; (2) they can significantly reduce the number of parameters without degrading  performance like the cross-layer parameter sharing. Particularly, we find that the recommendation quality  is always slightly better than the base model by the adjacent-block parameter sharing.

\subsection{General Architecture of CpRec}
By instantiating CpRec using the NextItNet architecture, we show the overal neural network architecture in Figure~\ref{fig:CpRec}. 
\subsubsection{Input layer} Given a user-item interaction sequence $\{x_{1}, x_{2},$ $ ..., x_{t+1}\}$, the recommender model  retrieves the embeddings of the first $t$ items $\left\{x_{1}, x_{2}, \ldots, x_{t}\right\}$  via look-up table based on  the block-wise embedding. After the dimensional projection, we can
 stack these item embeddings into a new matrix (as shown in the left part of Figure \ref{fig:CpRec}, where $t=5$), which serves as  inputs for the middle layers. 

\subsubsection{Middle layers} 
As shown in the middle part of Figure \ref{fig:CpRec}, we use the dilated convolutional layers~\cite{yuan2019simple} for CpRec, where every two layers are wrapped by a residual block structure.   CpRec obtains an exponential growth in the size of the receptive field by doubling the dilation for every layer, e.g., $\{1, 2, 4, 8\}$. 
%
In addition,  it is a common approach to repeat this structure several times to further enhance  model expressiveness and improve  accuracy, e.g.,  $\{1, 2, 4, 8, \ldots, 1, 2, 4, 8 \}$.  Then, we can
 apply the proposed layer-wise parameter sharing scheme  on these middle layers  to improve their parameter efficiency.

\subsubsection{Softmax layer}
The softmax layer  adopts the block-wise embedding decomposition with a tree structure so as to represent blocks of each cluster. As described before, for both the training and inference phases, CpRec can achieve significant speed-up by such  a structure.  
Following NextItNet, given each input sequence $\{x_{1}, x_{2},$ $ ..., x_{t}\}$, CpRec estimates the probability distribution of the output sequence  representing $\{x_{2}, x_{3},$ $ ..., x_{t+1}\}$, where $x_{t+1}$ is the next item expected.

\begin{table}[]
\small
	\caption{\small Statistic of the evaluated datasets. "M" and "K" is short for million and kilo, "t" is the length of interaction sequences. For ColdRec, the left  and right values devided by `/' denote  the source and target dataset, respectively.}
	\setlength{\tabcolsep}{2.4mm}
		\begin{tabular}{lccccc}
			\hline
			Dataset        & \#items & \#actions & \#sequences & t \\ \hline
			Weishi         & 66K   & 10M       & 1048575   & 10        \\
			ML20  & 54K   & 27.7M     & 1491478   & 20   \\
			TikTok         & 514K  & 38.2M     & 1018155   & 50  \\
			ML100 & 54K   & 27.7M     & 457350    & 100   \\
			ColdRec       & 191K/21K    & 82.5M/3.8M          & 1649095/3798114    & 50/1$\sim$3 \\ \hline
		\end{tabular}
\label{table:datadescribe}
\vspace{-0.1in}
\end{table}

\section{EXPERIMENTAL SETUP}
\label{experimentsetup}
\begin{table*}[]
\caption{\small Overall performance comparison, including recommendation accuracy, parameter efficiency (Params) , training time  and inference speedup (evaluated by the generation of top-5 items).
	  We omit the Params, Training Time (min) and Inference Speedup for {GRU4Rec} and {Caser} since they are not comparable to CpRec. MostPop returns item lists ranked by popularity. 
  	 {CpRec} with cross-layer~\cite{lan2019albert}, cross-block, adjacent-layer and adjacent-block parameter sharing is referred to {CpRec-Cl}, {CpRec-Cb}, {CpRec-Al} and {CpRec-Ab} respectively.
   }
\label{table:overallresult}
\small
\begin{tabular}{ccccccccccc}
\hline
\multicolumn{1}{c|}{\multirow{1}{*}{Data}}                                                  & Model     & MRR@5  & MRR@20 & HR@5   & HR@20  & NDCG@5 & NDCG@20 & Training Time (min) & Inference Speedup & Params \\ \hline
\multicolumn{1}{c|}{\multirow{7}{*}{Weishi}}          & MostPop & 0.0050 & 0.0121 & 0.0187 & 0.0940 & 0.0083 & 0.0294  & \verb|\|              &  \verb|\| &  \verb|\|     \\
\multicolumn{1}{c|}{}                                 & GRU4Rec & 0.1008 & 0.1159 & 0.1654 & 0.3222 & 0.1168 & 0.1610  & \verb|\|              &  \verb|\| &  \verb|\|     \\
\multicolumn{1}{c|}{}                                 & Caser & 0.0910 & 0.1047 & 0.1497 & 0.2937 & 0.1055 & 0.1460  & \verb|\|       & \verb|\| & \verb|\|     \\
\multicolumn{1}{c|}{}                                 & NextItNet & 0.1053 & 0.1205 & 0.1721 & 0.3302 & 0.1218 & 0.1665  & 27 & 1x & 74M      \\ \cline{2-11}
\multicolumn{1}{c|}{}                                 & CpRec-Cl & 0.1043	& 0.1193 & 0.1705 & 0.3264 & 0.1206 & 0.1646 & 12 & 2.8x & 9M      \\
\multicolumn{1}{c|}{}                                 & CpRec-Cb & \textbf{0.1071}	& 0.1222 & 0.1736 & 0.3304 & \textbf{0.1236} & 0.1678 & 12 & 2.8x & 9M      \\
\multicolumn{1}{c|}{}                                 & CpRec-Al & 0.1056	& 0.1207 & 0.1719 & 0.3295 & 0.1220 & 0.1665 & 12 & 2.8x & 11M      \\
\multicolumn{1}{c|}{}                                 & CpRec-Ab & 0.1069	& \textbf{0.1224} & \textbf{0.1742} & \textbf{0.3320} & \textbf{0.1236} & \textbf{0.1681} &  12 & 2.8x & 11M      \\ \hline
\multicolumn{1}{c|}{\multirow{7}{*}{ML20}}  & MostPop & 0.0044 & 0.0076 & 0.0134 & 0.0485 & 0.0067 & 0.0163  &  \verb|\|            &  \verb|\| &  \verb|\|     \\
\multicolumn{1}{c|}{}                                 & GRU4Rec & 0.0946 & 0.1095 & 0.1583 & 0.3122 & 0.1104 & 0.1539  &  \verb|\|              &  \verb|\| &  \verb|\|     \\
\multicolumn{1}{c|}{}                                 & Caser & 0.0922 & 0.1051 & 0.1511 & 0.2841 & 0.1068 & 0.1444  &  \verb|\|              &  \verb|\| &  \verb|\|     \\ 
\multicolumn{1}{c|}{}                                 & NextItNet & 0.1057 & 0.1212 & 0.1755 & 0.3352 & 0.1230 & 0.1682 & 104 & 1x & 62M      \\ \cline{2-11}
\multicolumn{1}{c|}{}                                 & CpRec-Cl & 0.1011 & 0.1162 & 0.1689 & 0.3243 & 0.1179 & 0.1619 & 64 & 2.4x & 13M      \\
\multicolumn{1}{c|}{}                                 & CpRec-Cb & 0.1049 & 0.1202 & 0.1747 & 0.3329 & 0.1221 & 0.1669 & 64 & 2.4x & 14M      \\
\multicolumn{1}{c|}{}                                 & CpRec-Al & 0.1040 & 0.1192 & 0.1729 & 0.3309 & 0.1210 & 0.1657 & 64 & 2.4x & 15M      \\
\multicolumn{1}{c|}{}                                 & CpRec-Ab & \textbf{0.1063} & \textbf{0.1218} & \textbf{0.1765} & \textbf{0.3364} & \textbf{0.1236} & \textbf{0.1690} & 64 & 2.4x & 15M      \\ \hline
\multicolumn{1}{c|}{\multirow{7}{*}{TikTok}}          & MostPop & 0.0006 & 0.0015 & 0.0017 & 0.0103 & 0.0009 & 0.0034 &  \verb|\|              &  \verb|\| &  \verb|\|     \\
\multicolumn{1}{c|}{}                                 & GRU4Rec & 0.0110 & 0.0154 & 0.023 & 0.0708 & 0.0140 & 0.0273  &  \verb|\|              &  \verb|\| &  \verb|\|     \\
\multicolumn{1}{c|}{}                                 & Caser & 0.0074 & 0.0098 & 0.0146 & 0.0409 & 0.0091 & 0.0164  &  \verb|\|              &  \verb|\| &  \verb|\|     \\
\multicolumn{1}{c|}{}                                 & NextItNet & 0.0120 & 0.0164 & 0.0242 & 0.0715 & 0.0150 & 0.0282 & 300 & 1x & 539M      \\ \cline{2-11}
\multicolumn{1}{c|}{}                                 & CpRec-Cl & 0.0130 & 0.0173 & 0.0255 & 0.0732 & 0.0161 & 0.0292 & 64 & 9.4x & 66M      \\
\multicolumn{1}{c|}{}                                 & CpRec-Cb & \textbf{0.0133} & \textbf{0.0174} & \textbf{0.0269} & \textbf{0.0755} & \textbf{0.0166} & \textbf{0.0297} & 64 & 9.4x & 67M      \\
\multicolumn{1}{c|}{}                                 & CpRec-Al & 0.0130 & 0.0172 & 0.0262 & 0.0741 & 0.0163 & 0.0289 & 64 & 9.4x & 72M      \\
\multicolumn{1}{c|}{}                                 & CpRec-Ab & 0.0129 & \textbf{0.0174} & 0.0254 & 0.0744 & 0.0158 & 0.0295 & 64 & 9.4x & 72M      \\ \hline
\multicolumn{1}{c|}{\multirow{7}{*}{ML100}} & MostPop & 0.0040 & 0.0068 & 0.0124 & 0.0433 & 0.0061 & 0.0146 & \verb|\| & \verb|\| &  \verb|\|     \\
\multicolumn{1}{c|}{}                                 & GRU4Rec & 0.0986 & 0.1131 & 0.1622 & 0.3123 & 0.1144 & 0.1568 & \verb|\| & \verb|\| &  \verb|\|     \\
\multicolumn{1}{c|}{}                                 & Caser & 0.0941 & 0.1072 & 0.1549 & 0.2902 & 0.1091 & 0.1474  &  \verb|\|      &  \verb|\| &  \verb|\|     \\
\multicolumn{1}{c|}{}                                 & NextItNet & 0.1090 & 0.1247 & 0.1781 & 0.3402 & 0.1261 & 0.1720 & 489 & 1x & 80M      \\ \cline{2-11}
\multicolumn{1}{c|}{}                                 & CpRec-Cl & 0.1020 & 0.1169 & 0.1674 & 0.3194 & 0.1184 & 0.1618 & 400 & 1.5x & 13M      \\
\multicolumn{1}{c|}{}                                 & CpRec-Cb & 0.1070 & 0.1221 & 0.1761 & 0.3325 & 0.1241 & 0.1683 & 400 & 1.5x & 13M      \\
\multicolumn{1}{c|}{}                                 & CpRec-Al & 0.1076 & 0.1227 & 0.1760 & 0.3335 & 0.1248 & 0.1690 & 400 & 1.5x & 25M      \\
\multicolumn{1}{c|}{}                                 & CpRec-Ab & \textbf{0.1111} & \textbf{0.1265} & \textbf{0.1819} & \textbf{0.3405} & \textbf{0.1286} & \textbf{0.1736} & 400 & 1.5x & 25M      \\ \hline
\end{tabular}
\vspace{-0.1in}
\end{table*}
This section introduces our experimental setup including datasets, baseline models, our implementation details and evaluation metrics.

\subsection{Datasets}
\begin{itemize}
	\item{\textbf{Movielens\footnote{https://grouplens.org/datasets/movielens/}}}: 
	The original dataset contains about 280,000 users, 58,000 videos  and 27 million user-item interactions with timesteps.
	To alleviate the impact of cold users and items, we perform the basic pre-processing by filtering out  interactions with less than 5 users and users with less than  10 items.
	Then, we define the maximum length of the interaction sequence as $t$, and  split  sequences that exceed the maximum length into multiple sub-sequences. Sequences shorter than $t$ will be padded with zero in the beginning of the sequence to reach $t$, following~\cite{yuan2019simple}. In this paper, we set $t$ to 20 and 100 as  short- and long-range sequences respectively, namely, ML20 and ML100.
	
	\item{\textbf{TikTok\footnote{https://www.tiktok.com/en/}}}: This dataset is released in ICME 2019 short video understanding challenge. It targets at predicting the next preferred videos for a user according to the historical watching behaviors. The original dataset is very large since it contains a large number of cold users and items.
	We follow the same procedure as in Movielens to remove cold users and items.  The dataset after basic pre-processing
	  better fits our GPU memory and helps speed up our experiments. 
	\item{\textbf{Weishi\footnote{https://weishi.qq.com}}}: Weishi is a private short-video recommendation dataset collected by Tencent (China). It containing more than 60,000 videos and we set $t$ to 10. 
	\item{\textbf{ColdRec}\cite{yuan2020parameter}}: This dataset is used to investigate the transfer learning task based on user interaction sequence. It contains a source  dataset and a target dataset. The source dataset contains a userID and his interaction sequence in the QQ Brower\footnote{https://brower.qq.com} recommender system, whereas the target dataset includes the same users and with less than 3 interactions in another recommender system --- Tencent Kandian\footnote{https://sdi.3g.qq.com/v/201911102006011550}.  ColdRec is used as a cold-user based recommendation task by performing transfer learning on the sequential recommender model (i.e., NextItNet).
\end{itemize}
Table \ref{table:datadescribe} summarizes the statistics of evaluated datasets after basic pre-processing in this work.

\subsection{Baseline model}
We compare CpRec with two typical sequential recommender models, namely {GRU4Rec} \cite{hidasi2015session} \& {Caser}~\cite{tang2018personalized} and one state-of-the-art model NextItNet. It needs to be noted that we use train GRU4Rec autoregressively, which has a similar effect as the data augmentation method in~\cite{tan2016improved}. Particularly, we perform extensive ablation studies by comparing with NextItNet since they have similar neural network architecture. All models are trained by using the cross-entropy loss~\cite{yuan2019simple}. 

\subsection{Implementation details}
We train all models using Tensorflow and Adam optimizer with a learning rate $1 \times 10^{-3}$ on GPU (Tesla P100). 
For all shared hyper-parameters, CpRec uses exactly the same  as NextItNet for comparison.  Regularization and dropout are not applied\footnote{While we observe that CpRec and NextItNet may perform further better by carefully tuning the regularization in our later work, all conclusions made in this paper hold fully consistent with their regularization variants.} following the official implementation of NextItNet\footnote{https://github.com/fajieyuan/nextitnet}. 
Specifically, on Weishi, we use a batch size (denoted by $b$) of 128 and dilation factors (denoted by $l$) $2\times\{1, 2, 2, 4\}$ (8 layers or 4 residual blocks). On ML20 and ML100, we set $b$ to 128, and $l$ to  $2\times\{1, 2, 4, 8\}$ (8 layers) and $8\times\{1, 2, 4, 8\}$ (32 layers), respectively. On TikTok, we set $b$ to 32 and $l$ to $4\times\{1, 2, 4, 8\}$  (16 layers).  
The embedding size $d$ on all above datasets is set to $512$. 
For GRU4Rec and Caser, we use the same embedding and hidden dimensions. Other specific hyper-parameters are empirically tuned according to the 
results on the testing set.
For the transfer learning task on ColdRec, we set  $b$ to 64 and 512 on the pre-trained and fine-tuned models, respectively. $l$ and $d$ are set to $4\times\{1, 2, 4, 8\}$ and 256, respectively. The hyper-parameter settings on ColdRec strictly follow~\cite{yuan2020parameter}.

For CpRec, the model-specific  hyper-parameters are the  cluster number $n$, partition rules, and the embedding size of each created block. 
Empirically, $n$ is very easy to be tuned. For example, If
we set $n$ to 2, then we can  partition  $S$ into two clusters and guarantee that $k_1: (K-k_1) \approx 2:8$ due to the 20/80 principle~\cite{jiang2013head} of the long-tailed distribution. This usually performs well and one can also  tune the ratio of  $k_1: (K-k_1) $ greedily for a further better result. If we set $n$ to 3, we first partition it into two clusters and then partition the second cluster and guarantee that $k_2: (K-k_1-k_2) \approx 2:8$.  Following this way, one can obtain the maximum compression ratio by fine  tuning $n$. In this paper, we set $n$ to 5 on TikTok and $3$ for the remaining datasets for evaluation purpose. The embedding size of each block can be set in the form of a geometric  progression, such as \{512, 256, 128\}.

\begin{table}[]
\caption{\small Performance comparison w.r.t. how to apply the block-wise embedding decomposition.  {NextItNet} that uses block-wise decomposition  in the input layer,  output layer  and  both are referred to {Bi-NextItNet}, {Bo-NextItNet} and {Bio-NextItNet}, respectively. 
  {B1-NextItNet} employs the standard low-rank decomposition  (i.e., with only 1 block) in the input and softmax layer inspired by~\cite{lan2019albert}. 
  Note that for clarity only the parameters in the input and output matrices 
  are reported in the Params Column. TT is short for training time (unit: min). The inference speedup is simply omitted due to similar results as in Table ~\ref{table:overallresult}. 
}
\label{table:ablationblock}
\small
\setlength{\tabcolsep}{2mm}
\begin{tabular}{cccccccccc}
\hline
Data & Model     & MRR@5  & HR@5   & TT & Params  \\ \hline
\multicolumn{1}{c|}{\multirow{4}{*}{Weishi}}          & NextItNet & 0.1053 & 0.1721 & 27 & 68M     \\
\multicolumn{1}{c|}{}                                 & B1-NextItNet & 0.0959 & 0.1582 & 18 & 17M    \\
\multicolumn{1}{c|}{}                                 & Bi-NextItNet & 0.1059 & 0.1724 & 25 &  38M    \\
\multicolumn{1}{c|}{}                                 & Bo-NextItNet & 0.1060 & 0.1729 & 14 &  38M  \\
\multicolumn{1}{c|}{}                                 & Bio-NextItNet & \textbf{0.1068} & \textbf{0.1734} & 12 &  8M    \\ \hline
\multicolumn{1}{c|}{\multirow{4}{*}{ML20}}  & NextItNet & 0.1057 & 0.1755 & 104 &  55M  \\
\multicolumn{1}{c|}{}                                 & B1-NextItNet & 0.0958 & 0.1603 & 79 &  14M    \\
\multicolumn{1}{c|}{}                                 & Bi-NextItNet & 0.1058 & 0.1761 & 102 &  34M  \\
\multicolumn{1}{c|}{}                                 & Bo-NextItNet & 0.1063 &\textbf{0.1766} & 66 &  34M  \\
\multicolumn{1}{c|}{}                                 & Bio-NextItNet & \textbf{0.1064} & \textbf{0.1766} & 64 &  12M   \\  \hline
\multicolumn{1}{c|}{\multirow{4}{*}{TikTok}}          & NextItNet & 0.0120 & 0.0242 & 300 &  527M \\
\multicolumn{1}{c|}{}                                 & B1-NextItNet & 0.0109 & 0.0226 & 162 &  132M   \\
\multicolumn{1}{c|}{}                                 & Bi-NextItNet & 0.0117 & 0.0239 & 278 &  296M  \\
\multicolumn{1}{c|}{}                                 & Bo-NextItNet & 0.0122 & 0.0249 & 78 &  296M \\
\multicolumn{1}{c|}{}                                 & Bio-NextItNet & \textbf{0.0123} & \textbf{0.0250} & 64 &  66M  \\ \hline
\multicolumn{1}{c|}{\multirow{4}{*}{ML100}} & NextItNet & 0.1090 & 0.1781 & 489 &  55M \\
\multicolumn{1}{c|}{}                                 & B1-NextItNet & 0.0917 & 0.1619 & 439 &  14M   \\
\multicolumn{1}{c|}{}                                 & Bi-NextItNet & 0.1092 & 0.1791 & 487 &  34M \\
\multicolumn{1}{c|}{}                                 & Bo-NextItNet & 0.1090 & 0.1780 & 402 &  34M \\
\multicolumn{1}{c|}{}                                 & Bio-NextItNet & \textbf{0.1109} & \textbf{0.1818} & 400 &  12M \\ \hline
\end{tabular}
\vspace{-0.1in}
\end{table}

\subsection{Evaluation Metrics}
In order to evaluate the recommendation accuracy of CpRec, we  randomly split all datasets into training (80\%) and testing (20\%) sets. Following previous works \cite{hidasi2015session, he2020lightgcn}, we use the popular top-N metrics, including MRR@N (Mean Reciprocal Rank), HR@N (Hit Ratio) and NDCG@N (Normalized Discounted Cumulative Gain), where N is set to 5 and 20.  To evaluate the parameter efficiency, we report
 the total number of parameters (without special mention) of each model by {Params}. To reflect the training efficiency, we report the training time for each model until convergence, denoted by {Training Time (min)}. The inference speedup compared to the baseline is also reported. Similarly to \cite{yuan2019simple,kang2018self}, we only evaluate the prediction accuracy of the last item in each interaction sequence in testing set.

\section{EXPERIMENTAL RESULTS}
The key contribution of CpRec is to improve the memory efficiency for sequential recommender models based on deep neural networks. In this section, we answer the following research questions:
\begin{itemize}
\item[(1)] \textbf{RQ1}: Does CpRec significantly reduce the model size of a typical sequential neural network, i.e., NextItNet in this paper? If so, does it perform comparably to NextItNet in terms of recommendation accuracy? Are there other advantages that CpRec has over  NextItNet?

\item[(2)] \textbf{RQ2}: What impacts (effectiveness \& efficiency) does the adaptive decomposition and layer-wise parameter sharing methods have on CpRec?  
\item[(3)] \textbf{RQ3}: Is CpRec a generic framework that works well for other sequential recommender models, such as GRU4Rec?
\item[(4)] \textbf{RQ4}: Since sequential recommender models can also be applied for the pre-training and fine-tuning-based transfer learning task~\cite{yuan2020parameter}, does CpRec work as well as the non-compressed model for such a task?
\end{itemize}

\begin{table}[]
\caption{\small The impact of layer-wise parameter sharing strategies. NextItNet with cross-layer, cross-block, adjacent-layer and adjacent-block parameter sharing is denoted by {Cl-NextItNet}, {Cb-NextItNet}, {Al-NextItNet}, {Ab-NextItNet}, respectively.   Note for clarity  only the parameters in the middle layers are shown in the Params Column.}
\label{table:ablationlayer}
\small
\setlength{\tabcolsep}{3.5mm}
\begin{tabular}{ccccccccc}
\hline
Data & Model  & MRR@5  & HR@5  & Params \\ \hline
\multicolumn{1}{c|}{\multirow{5}{*}{Weishi}}          & NextItNet & 0.1053 & 0.1721 &  6M    \\
\multicolumn{1}{c|}{}                                 & Cl-NextItNet & 0.1021 & 0.1669 &  1M   \\
\multicolumn{1}{c|}{}                                 & Cb-NextItNet & 0.1047 & 0.1710 &  2M   \\
\multicolumn{1}{c|}{}                                 & Al-NextItNet & 0.1042 & 0.1699 &  3M  \\
\multicolumn{1}{c|}{}                                 & Ab-NextItNet & \textbf{0.1057} & \textbf{0.1727} &  3M    \\ \hline
\multicolumn{1}{c|}{\multirow{5}{*}{ML20}}  & NextItNet & 0.1057 & 0.1755 &  6M  \\
\multicolumn{1}{c|}{}                                 & Cl-NextItNet & 0.1006 & 0.1678 &  1M  \\
\multicolumn{1}{c|}{}                                 & Cb-NextItNet & 0.1037 & 0.1728 &  2M \\
\multicolumn{1}{c|}{}                                 & Al-NextItNet & 0.1037 & 0.1730 &  3M    \\
\multicolumn{1}{c|}{}                                 & Ab-NextItNet & \textbf{0.1062} & \textbf{0.1762} &  3M   \\ \hline
\multicolumn{1}{c|}{\multirow{5}{*}{TikTok}}          & NextItNet & 0.0120 & 0.0242 & 13M \\
\multicolumn{1}{c|}{}                                 & Cl-NextItNet & 0.0119 & 0.0241 &  1M  \\
\multicolumn{1}{c|}{}                                 & Cb-NextItNet & 0.0120 & 0.0243 &  2M  \\
\multicolumn{1}{c|}{}                                 & Al-NextItNet & 0.0118 & 0.0240 &  6M \\
\multicolumn{1}{c|}{}                                 & Ab-NextItNet & \textbf{0.0122} & \textbf{0.0247} & 6M  \\ \hline
\multicolumn{1}{c|}{\multirow{5}{*}{ML100}} & NextItNet & 0.1090 & 0.1781 &  25M \\
\multicolumn{1}{c|}{}                                 & Cl-NextItNet & 0.1007 & 0.1656 & 1M \\
\multicolumn{1}{c|}{}                                 & Cb-NextItNet & 0.1048 & 0.1735 &  2M \\
\multicolumn{1}{c|}{}                                 & Al-NextItNet & 0.1077 & 0.1762 &  13M \\
\multicolumn{1}{c|}{}                                 & Ab-NextItNet & \textbf{0.1101} & \textbf{0.1812} &  13M \\ \hline 
\end{tabular}
\vspace{-0.1in}
\end{table}

\subsection{Quantitative Evaluation (RQ1)}
We present the results of {CpRec} and the baseline models on the four sequential recommendation datasets in Table \ref{table:overallresult}. First, we find that NextItNet performs better than {GRU4Rec} and {Caser}  with notable improvements in the recommendation accuracy across all datasets. Our observation here is consistent with that in~\cite{yuan2019simple,tang2019towards,wang2019towards}. 

Second, CpRec with our proposed layer-wise parameter sharing methods (i.e., CpRec-Cb, CpRec-Al \& CpRec-Ab)  yields competitive results with NextItNet on all evaluation metrics. Particularly, CpRec-Ab performs consistently better than NextItNet --- e.g., on TikTok CpRec-Ab obtains 7.5\% improvements  in terms of MRR@5. Moreover, CpRec has obvious advantages over NextItNet in terms of both parameter efficiency and training/inference time. 
On the TikTok and Weishi datasets, the compression ratios are up to 7 to 8 times, respectively. By contrast, the state-of-the-art ranking distillation method proposed in ~\cite{tang2018ranking} achieved only around 2 times the compression ratio. Besides, CpRec is much faster in both training and generating relative to NextItNet, due to the block-wise embedding decomposition and the efficient tree structure in the softmax layer. It is important to note that the reduction of model size does not necessarily lead to a significant improvement in the training speed.  Unlike other datasets, the training speed on ML100 is not largely improved by CpRec. This is because on ML100 the time complexity is dominated by the middle layers (32 layers), though the parameters in the middle layers have been significantly reduced. 

\begin{figure}[t]
    \centering
    \begin{subfigure}[t]{0.25\textwidth}
    \centering
            \includegraphics[width=1.7in]{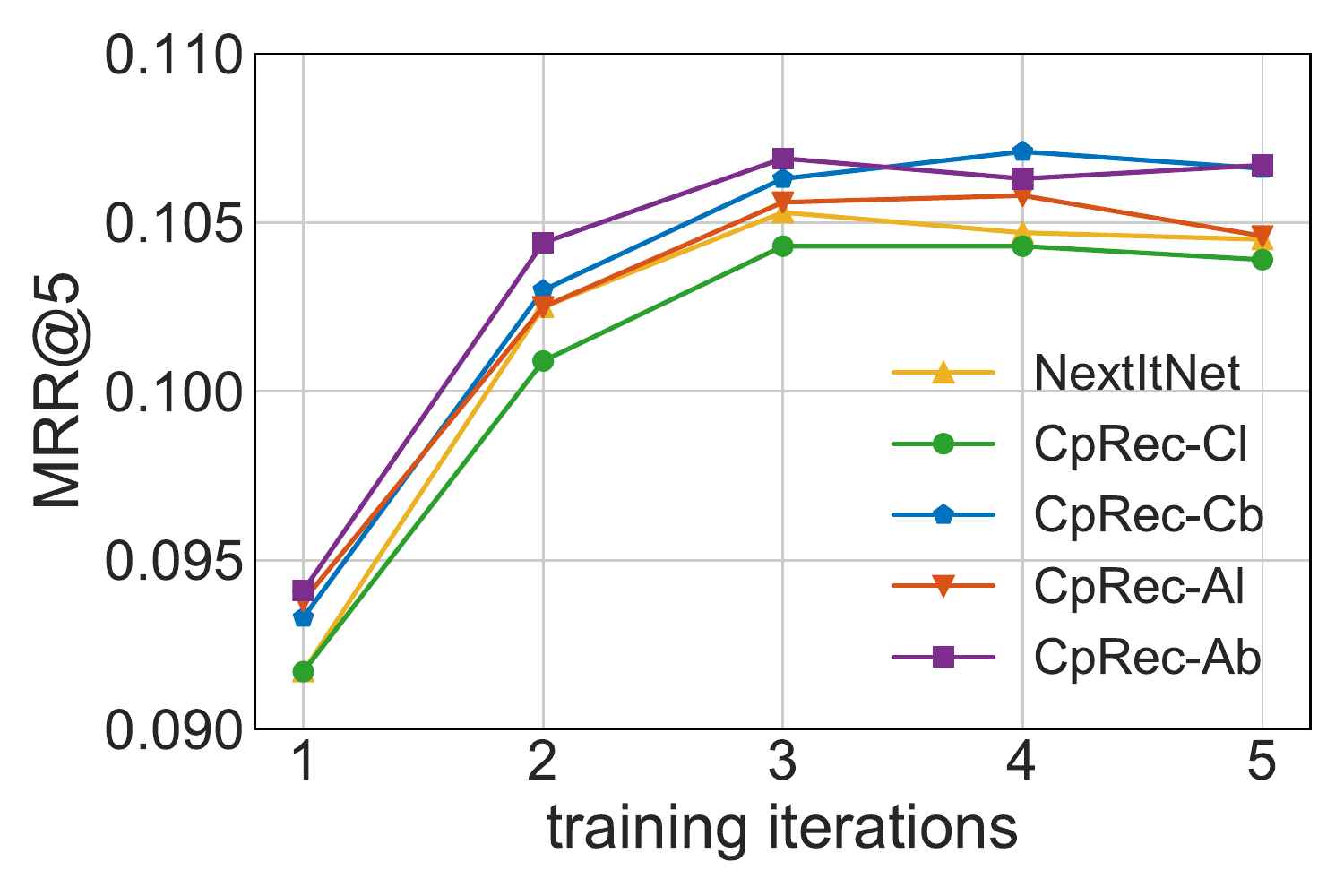}
            \subcaption{Weishi}
            \label{fig:weishimrr}
    \end{subfigure}%
    \begin{subfigure}[t]{0.25\textwidth}
    \centering
            \includegraphics[width=1.7in]{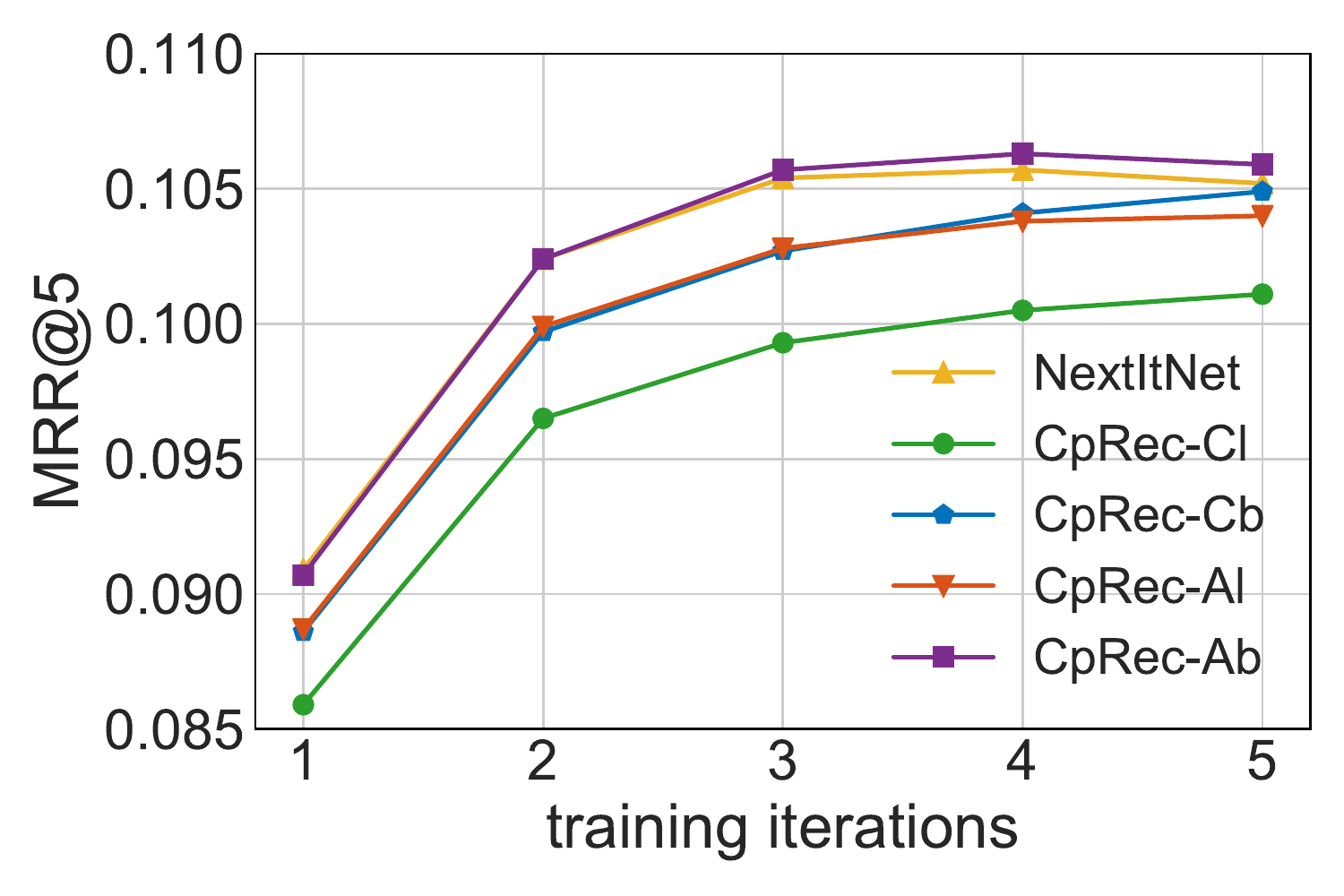}
            \subcaption{ML20}
            \label{fig:Movielens20mrr}
    \end{subfigure}
    \caption{\small Convergence behaviors on Weishi and ML20.}
    \label{fig:overallmrr}
\end{figure}

On the other hand, we observe that CpRec with cross-layer sharing scheme (i.e., CpRec-Cl) yields relatively worse results on ML20 and ML100. As we mentioned before,  CpRec-Cl with too few parameters may restrict its model expressiveness. This can be evidenced by CpRec-Cb since it always outperforms CpRec-Cl. Note the improvements of CpRec-Cl over NextItNet on TikTok mainly come from the block-wise embedding decomposition, which is verified by the further ablation studies in Table~\ref{table:ablationblock}. To show the convergence behaviors of CpRec, we plot the results in Figure~\ref{fig:overallmrr}, which gives similar observations as above.

\begin{table}[]
\small
\setlength{\tabcolsep}{2.5mm}
\caption{\small The effect of adaptive embedding decomposition applied to GRU4Rec. TT is short for training time (unit: min).}
\label{table:blockGRU}
\begin{tabular}{cccccccccc}
\hline
Data & Model     & MRR@5  & HR@5   & TT & Params \\ \hline
\multicolumn{1}{c|}{\multirow{4}{*}{Weishi}}          & GRU4Rec & 0.1008 & 0.1654 & 21 & 69M     \\
\multicolumn{1}{c|}{}                                 & Bi-GRU4Rec & 0.1015 & 0.1666 & 19 & 39M    \\
\multicolumn{1}{c|}{}                                 & Bo-GRU4Rec & 0.1039 & 0.1705 & 7 & 39M    \\
\multicolumn{1}{c|}{}                                 & Bio-GRU4Rec & \textbf{0.1045} & \textbf{0.1711} & 6 & 9M     \\ \hline
\multicolumn{1}{c|}{\multirow{4}{*}{ML100}}  & GRU4Rec & 0.0986 & 0.1622 & 201 & 57M   \\
\multicolumn{1}{c|}{}                                 & Bi-GRU4Rec & 0.0988 & 0.1640 & 199 & 35M   \\
\multicolumn{1}{c|}{}                                 & Bo-GRU4Rec & 0.0995 & 0.1627 & 74 & 35M   \\
\multicolumn{1}{c|}{}                                 & Bio-GRU4Rec & \textbf{0.1025} & \textbf{0.1683} & 72 & 14M     \\  \hline
\end{tabular}
\vspace{-0.1in}
\end{table}

\subsection{Ablation Study (RQ2)}
In order to evaluate the impacts of different components on CpRec, we report the ablation test  from two aspects: (1) the impact of block-wise adaptive decomposition on the input layer,  the output layer,  and both. (2) the impact of layer-wise parameter sharing strategies.
\subsubsection{Block-wise adaptive decomposition}
Table \ref{table:ablationblock} shows the results of block-wise embedding decomposition with 3 different settings as mentioned above. 
The experimental results clearly demonstrate that {Bi}-  \& {Bo}-based {NextItNet} achieve similar improvements on parameter efficiency, compressing models to $57\% \sim 74\%$ of the original size. By combining {Bi}-  \& {Bo}-, we obtain  {Bio-NextItNet}, which gives the best compression ratio as well as the highest top-N results  and fastest training speed. In addition, we observe that B1-NextItNet performs much worse than NextItNet with around 5$\sim$8\% accuracy loss. By contrast,  {Bio-NextItNet} performs better than NextItNet \&  B1-NextItNet although  Bio-NextItNet is more compact than B1-NextItNet. These results suggest that the proposed adaptive decomposition is more effective than the standard decomposition.


%

\subsubsection{Layer-wise parameter sharing}
 Unlike the  block-wise embedding decomposition, layer-wise parameter sharing approaches focus on reducing parameters of  the middle layers, and thus are only useful for deep neural network based recommender models. 
 Table~\ref{table:ablationlayer} presents experimental results for the four layer-wise parameter sharing strategies. Clearly, the results indicate that the proposed adjacent-block strategy (i.e., Ab-NextItNet) always yields 
 the best recommendation accuracy, and even outperforms the standard NextItNet. In addition, it  reduces the parameters to half of the original size. 
 Among these approaches, the cross-layer strategy (i.e., Cl-NextItNet) gains the best parameter efficiency, but significantly
 hurts the model accuracy. By contrast, the cross-block strategy (i.e., Cb-NextItNet) seems also a worth
 trade-off given its slightly inferior  accuracy but a good compression rate. So it may depend on the practical scenario when determining which of the three proposed approaches to use.

\

\subsection{Adaptability Experiment (RQ3)}
In order to verify the adaptability of CpRec on other sequential recommender models, we specify CpRec with GRU4Rec and report results  in Table~ \ref{table:blockGRU}. Similar behaviors have been observed by other datasets, which however are omitted for saving space. Note that since we observe that using more hidden layers does not improve the accuracy of GRU4Rec, we only investigate the block-wise adaptive
decomposition for it.
Similar to the above experiments, we prefix the model name with Bi-, Bo- and Bio-. As expected, Bio-GRU4Rec obtains the best accuracy, compression rate and training speed. The results can well evidence the applicability and generality of our CpRec.

\begin{table}[]
\small
\caption{\small CpRec vs. NextItNet on the transfer learning task. Note that our evaluation strictly follows~\cite{yuan2020parameter}. MRR@5  \& HR@5 are the finetuned accuracy, whereas Params and training time are evaluated on the pre-trained model, which is computationally more expensive than the finetuned model.}
\setlength{\tabcolsep}{2.2mm}
\scalebox{1}{
\begin{tabular}{ccccc}
\hline
Model          & MRR@5  & HR@5   & Params & Training Time (min) \\ \hline
NextItNet      & 0.2012 & 0.3496 & 102M    & 578                 \\
CpRec-Cb & \textbf{0.2035} & \textbf{0.3540}  & 31.5M   & 118                 \\
CpRec-Al & 0.2022 & 0.3507 & 32.7M   & 118                 \\
CpRec-Ab & 0.2025 & 0.3517 & 32.7M   & 118                 \\ \hline
\end{tabular}
}
\vspace{-0.1in}
\label{table:transfer}
\end{table}

\subsection{Transfer Learning Experiment (RQ4)}
Deep learning based sequential recommender models can not only recommending items from where they come from, but also work as a knowledge transfer learning tool to improve recommendation quality in other systems.  In other words, we first use CpRec as a pre-trained model and fully train it using the source dataset of ColdRec. Then, we simply  add a new softmax layer on the final hidden layer of CpRec, and finetune all parameters on
the target dataset by using the pre-trained weights as a warm start.
Our transfer learning framework strictly follows a recent work in ~\cite{yuan2020parameter}. We report the results in Table \ref{table:transfer}.

As shown, CpRec obtains more than 3 times compression rate on ColdRec, but consistently outperforms  NextItNet with all  proposed layer-wise parameter sharing methods. The HR@5 result of NextItNet reported here  is exactly the same as that in
the original paper (i.e., FineAll in Table 2 in ~\cite{yuan2020parameter}).  In addition, the pretraining time of CpRec is also several times faster than NextItNet,  similar to the next item recommendation task.


\section{Conclusions}
In this paper, we have proposed CpRec, a flexible  \& generic neural network compression framework for learning compact sequential recommender models. CpRec significantly reduces parameter size in both the input and softmax layer by 
leveraging the inherent long-tailed item distribution. Moreover, CpRec performs further compression by a series of  layer-wise parameter sharing methods. Through extensive experiments on real-world datasets, we show that CpRec generates recommendations with higher speed, lower memory and often better accuracy. An important conclusion made from these results is that the commonly used recommender models are not compact at all. Hence, we expect CpRec to be valuable for existing SRS based on deep neural networks.

\bibliographystyle{ACM-Reference-Format}
\bibliography{sample-base}




\end{document}